\documentclass[aps,prl,showpacs,amssymb,nofootinbib,superscriptaddress,twocolumn,longbibliography]{revtex4-1}
\usepackage{tikz}
\usetikzlibrary{shapes,snakes,backgrounds,fit,decorations.pathreplacing}
\usepackage{bbm,times}
\usepackage{bm}
\usepackage{amsbsy}
\usepackage{amsthm}
\usepackage{amssymb}
\usepackage{amsfonts}
\usepackage{amsmath}
\usepackage[c]{esvect} %for nice vectors. Command for vectors becomes \vec{}
\usepackage{dsfont} % for symbols like the identity: \mathds{1}\right 
\usepackage{graphicx} % for graphics
\usepackage{epsfig}
\usepackage{epstopdf}
\usepackage{dsfont}
\usepackage{multibib}
\usepackage{color}
\usepackage{enumerate}
\usepackage{multirow}

\usepackage[colorlinks]{hyperref}
\makeatletter
\newcommand\org@hypertarget{}
\let\org@hypertarget\hypertarget
\renewcommand\hypertarget[2]{%
	\Hy@raisedlink{\org@hypertarget{#1}{}}#2%
}
\makeatother
\usepackage[figure,table]{hypcap}
\usepackage{MnSymbol}
\usepackage{enumerate}%allows different styles of enumerate environment
\usepackage{float}
\hypersetup{
	bookmarksnumbered,   
	pdfstartview={FitH},
	citecolor={darkblue},
	linkcolor={darkred},
	urlcolor={darkblue},
	pdfpagemode={UseOutlines}}
\definecolor{darkgreen}{RGB}{50,190,50}
\definecolor{darkblue}{RGB}{0,0,190}
\definecolor{darkred}{RGB}{238,0,0}
\usepackage{soul}

\makeatletter
\renewcommand{\p@subsection}{}
\renewcommand{\p@subsubsection}{}
\makeatother
\begin{document}
\title{Natural and magnetically induced entanglement of  hyperfine-structure states in atomic hydrogen}

\author{ Yusef Maleki,$^{1}$
%\thanks{E-mail: maleki@physics.tamu.edu}
 Sergei Sheludiakov,$^{1}$ Vladimir V. Khmelenko,$^{1}$ Marlan O. Scully,$^{1,2}$ David M. Lee,$^{1}$ and Aleksei M. Zheltikov$^{1,3,4,5}$ 
\\ $^1${\small Institute of Quantum Science and Engineering, Department of Physics and Astronomy, Texas A\&M University, College Station, Texas 77843-4242, USA}
\\ $^2${\small Princeton University, Princeton, NJ 08544, USA}
\\ $^3${\small Physics Department, M.V. Lomonosov Moscow State University, Moscow 119992, Russia}
\\ $^4${\small Advanced Photonics Laboratory, Russian Quantum Center, Skolkovo, Moscow Region 143025, Russia}
\\ $^5${\small Kazan Quantum Center, A.N. Tupolev Kazan National Research Technical University, 420126 Kazan, Russia}}

\date{\today}

\begin{abstract}
The spectrum of atomic hydrogen has long been viewed as a Rosetta stone that bears the key to decode the writings of quantum mechanics in a vast variety of physical, chemical, and biological systems. Here, we show that, in addition to its role as a basic model of quantum mechanics, the hydrogen atom provides a fundamental building block of quantum information. Through its electron and nuclear spin degrees of freedom, the hydrogen atom is shown to lend a physically meaningful frame and a suitable Hilbert space for bipartite entanglement, whose two-qubit concurrence and quantum coherence can be expressed in terms of the fundamental physical constants -- the Planck and Boltzmann constants, electron and proton masses, the fine-structure constant, as well as the Bohr radius and the Bohr magneton. The intrinsic, natural entanglement that the hyperfine-structure (HFS) states of the H atom store at low temperatures rapidly decreases with a growth in temperature, vanishing above a $\tau_c$ $\approx$ 5.35 $\mu$eV threshold. An external magnetic field, however, can overcome this thermal loss of HFS entanglement. As one of the central findings of this work, we show that an external magnetic field can induce and sustain an HFS entanglement, against all the odds of thermal effects, at temperatures well above the  $\tau_c$ threshold, thus enabling magnetic-field-assisted entanglement engineering in low-temperature gases and solids.

\end{abstract}

\maketitle

\section{Introduction}
In its capacity as the simplest of atoms, atomic hydrogen plays a central role in physics, serving as an ultimate reference point, a basic model of a bound quantum system, and a source of closed-form analytical solutions for quantum theories \cite{Bohr,Bethe}. The significance of the H atom for the development of quantum concepts is well-documented in the canonical accounts of the early era of quantum mechanics \cite{Bethe,Series,Landau}. In a more recent history, atomic hydrogen has continued to stay one of the most extensively studied objects of fundamental \cite{Hansch72,Udem,Fischer} and applied \cite{Marowsky,Gladkov} spectroscopy, as well as attosecond physics \cite{Sansone,attoH2,attoH3}, serving as the basis for understanding the structure of matter and a highly sensitive probe for experimental tests of the quantum theory \cite{LeeRMP,Hansch2006}.

As a fundamental connection to solid-state and low-temperature physics, recent studies reveal nuclear-polarized phases of hydrogen atoms embedded in solid H$_2$ films \cite{Bigelow,Sheludiakov} with remarkably large deviations of the low-temperature nuclear spin polarization of H atoms from the Boltzmann distribution \cite{Sheludiakov,Ahokas,Ahokas2}. This finding raises a number of interesting and important questions regarding the role of quantum effects in systems of this class, including the feasibility of quantum entanglement of H at thermal equilibrium. Important insights into this question can be found in earlier studies on electron-spin dynamics in two-electron double-quantum-dot systems \cite{Johnson,Petta}, which have been shown to offer much promise as prospective qubits and advantageous building blocks for quantum
information technologies \cite{Taylor,Ladd,Bennett2}. 
In search of the insights into these questions, we are led to revisit the entanglement capacity of the basic building block of such a hypothetical quantum interface -- a single hydrogen atom. As we show below in this paper, the electron and nuclear spin degrees of freedom in a hydrogen atom provide both a physically meaningful frame and a suitable Hilbert space for bipartite entanglement. We quantify this entanglement in terms of a two-qubit concurrence and quantum coherence, revealing a connection of H-atom entanglement to the fundamental physical constants -- the Planck and Boltzmann constants, electron and proton masses, the fine-structure constant, as well as the Bohr radius and the Bohr magneton. The intrinsic, natural entanglement that the hyperfine-structure (HFS) states of the H atom store at low temperatures rapidly decreases with a growth in temperature, vanishing above a $\tau_c$ $\approx$ 5.35 $\mu$eV threshold. An external magnetic field, however, can overcome this thermal loss of HFS entanglement above $\tau_c$, enabling magnetic-field-assisted entanglement engineering in low-temperature gases and solids.

As one of the prominent earlier efforts to understand quantum-entanglement properties of the H atom, Tommassini et al. \cite{Tommassini} have  analyzed the electron--proton coordinate entanglement, understood as correlations in the electron--proton motion in the H atom. While the behavior of this electron--proton position entanglement as a function of temperature is yet to be understood, here, we focus on another type of quantum entanglement built into the H atom -- entanglement related to the nuclear and electron spin degrees of freedom. An illuminating analysis of entanglement of this type has been earlier presented by Zhu et al. \cite{Zhu}, who developed a helpful formalism for the description of electron--nuclear spin entanglement in the H atom and examined important properties of his entanglement. Entanglement involving nuclear degrees of freedom is in no way unique to the H atom. As one prominent example, nuclear spins in nitrogen--vacancy centers in diamond have been shown to offer a powerful resource for quantum information, attracting much interest in the context of rapidly growing quantum technologies \cite{Childress,Dutt,Neumann,Fuchs}.

As one of the central findings of our work, which goes well beyond the scope of Ref. \cite{Zhu}, we demonstrate that an external magnetic field can induce and sustain an HFS entanglement, against all the odds of thermal effects, at temperatures well above the  $\tau_c$ threshold, thus enabling magnetic-field-assisted entanglement engineering in low-temperature gases and solids. Since this in many ways counterintuitive behavior of entanglement in the H atom is due to the interaction of the electron and nuclear spins with an external magnetic field, this effect is not foreseen for the electron--proton coordinate entanglement, such as the one studied by Tommassini et al. \cite{Tommassini}. Moreover, we extend the analysis of the HFS entanglement in the H atom to show that the criterion of this entanglement can be meaningfully expressed in terms of a suitably defined quantum coherence, providing useful insights into the behavior of the HFS entanglement as a function of temperature and helping better understand the phenomenon of magnetically induced entanglement above the  $\tau_c$ threshold.

\section{THE HAMILTONIAN AND THE HYPERFINE
STRUCTURE OF H}
In our analysis of the HFS entanglement of the H atom, we resort to the standard equation for the spin part of its ground-state Hamiltonian in an external magnetic field $B$ \cite{Feynman},
\begin{align}\label{Hamilton}
   H_{HF}=\mathcal{A}(\sigma_e\cdot\sigma_p)+\mu_B \sigma_e\cdot B.
\end{align}
Here, $\sigma_q$ = ($\sigma_q^x$, $\sigma_q^y$, $\sigma_q^z$) is the vector composed of the Pauli operators $\sigma_j^x$, $\sigma_j^y$, and $\sigma_j^z$, $q=e$ and $p$ for the electron and proton, respectively, $\mu_B=e\hbar/2m_e c$ is the Bohr magneton,
$$
\mathcal{A}={A\hbar^2}/{4}=\left[\left(\frac{2\pi}{3}\right) \left(\frac{1}{4\pi \varepsilon _{0}}\right)  \left(\frac{\hbar^2}{c\pi a_0^3}\right) \left(\frac{g_e e}{2m_e}\right) \left(\frac{g_p e}{2m_p}\right)\right]
$$
is the HFS constant, $e$ is the electron charge, $m_e$ and $m_p$ are the electron and proton masses, $g_e=-\gamma_e\mu_B/\hbar$, $g_p=\gamma_p\mu_N/\hbar$, $\gamma_e$ and $\gamma_p$ are the electron and proton gyromagnetic ratios, $\mu_N=e\hbar/2m_p c$ is the nuclear magneton, $c$ is the speed of light in vacuum, and $a_0$ is the Bohr radius. 

The first term in Eq. (1) describes the interaction between the electron and nuclear spins, $\hat{S}$ and $\hat{I}$,   $H_1=A\hat{S}\cdot\hat{I}$, with $\mathcal{A}={A\hbar^2}/{4}$. The second term accounts for the interaction between the electron spin and the external magnetic field $B$, $H_2=-\gamma_e S\cdot B$. Interaction between the nuclear spin and $B$ is much weaker and is neglected.

The energy eigenvalues $E_u$ of the Hamiltonian (1) are found as solutions to $ H_{HF} |u\rangle=E_u  |u\rangle$, yielding \cite{Feynman,Breit} four energy eigenstates (Fig. 1a), $u = a, b, c, d$, with $E_{a,c} =\mathcal{A}(-1\mp 2\sqrt{1+\xi^2})$, $E_{b,d} = \mathcal{A}(1\mp2\xi)$, where $\xi=\mu_B B/(2\mathcal{A})$, the "--" sign is taken for $u = a$ and $b$ and the "+" sign is taken for $u = c$ and $d$.

 The eigenstates of the Hamiltonian (1) can now be represented as
 \begin{align}
\left(
  \begin{array}{c}
 |d\rangle \\
     |b\rangle  \\
    |c\rangle  \\
     |a\rangle \\
  \end{array}
\right)&=
\left(
  \begin{array}{cccc}
   1 &0& 0&0\\
      0 & 1 & 0& 0 \\
   0 &0& x_+ &y_+  \\
    0 & 0 & x_- &y_- \\
  \end{array}
  \right)
  \left(
  \begin{array}{c}
 |\uparrow_e \uparrow_p\rangle \\
     |\downarrow_e \downarrow_p\rangle \\
    |\uparrow_e \downarrow_p\rangle  \\
    |\downarrow_e \uparrow_p\rangle \\
  \end{array}
\right),
 \end{align}
where $x_\pm=\frac{\sqrt{1+\xi^2}\pm\xi}{\sqrt{1+(\sqrt{1+\xi^2}\pm\xi)^2}}$ and $y_\pm=\frac{\pm1}{\sqrt{1+(\sqrt{1+\xi^2}\pm\xi)^2}}$.

  \begin{figure}
\includegraphics[width=\columnwidth]{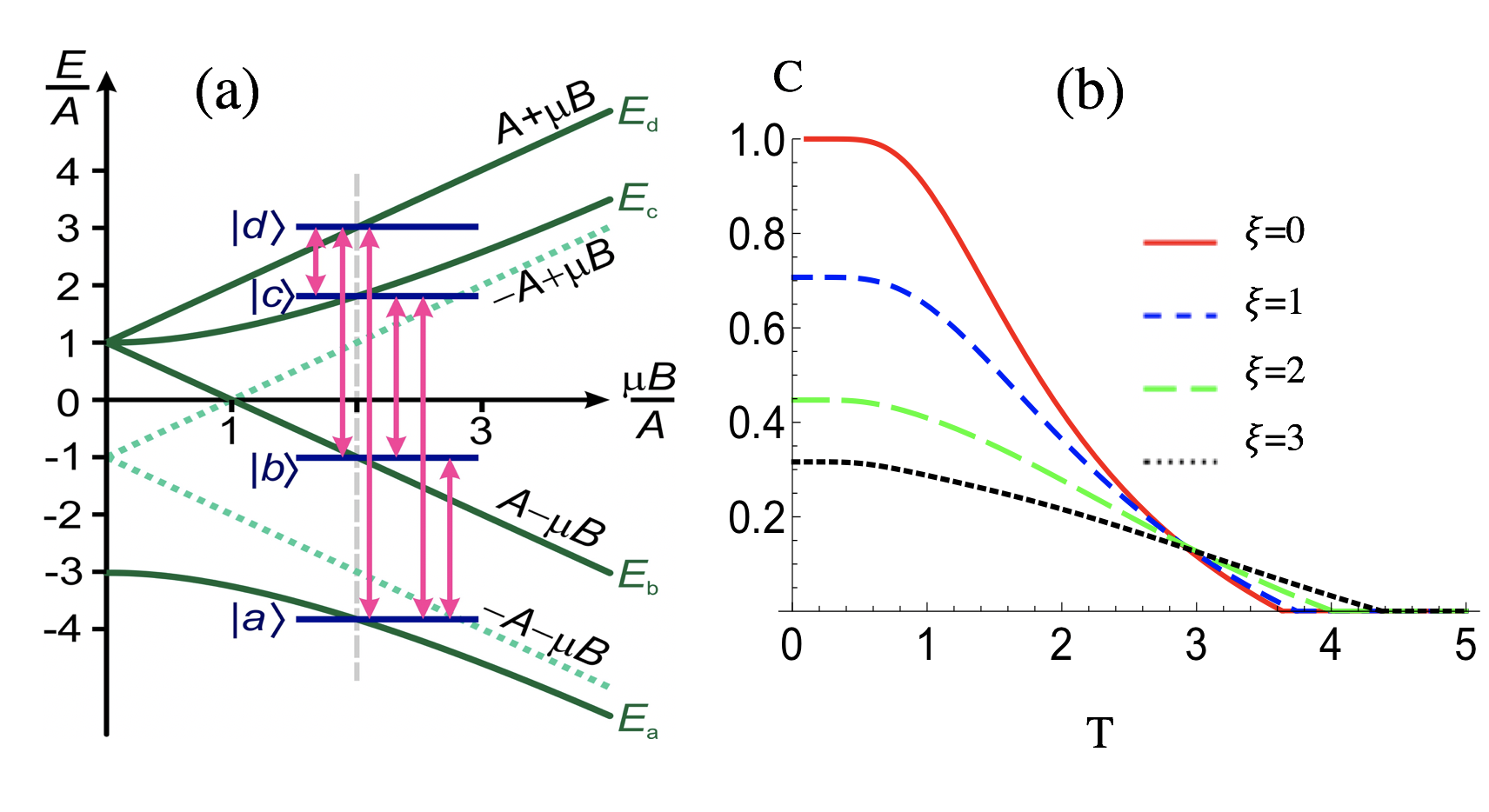}
\centering
\caption{(a) Energy level diagram of the hyperfine structure of a ground-state H atom driven by a magnetic field $B$. (b) The concurrence of the HFS states of the H atom as a function of the temperature $T$ for different values of the normalized magnetic field $\xi$.
}
\label{fig:2}
\end{figure}

To describe the entanglement of the HFS states of the hydrogen atom, we consider a four-dimensional Hilbert space with a basis 
$\mathcal{Q}(e,p) =$ $\{|\uparrow_e \uparrow_p\rangle,|\uparrow_e \downarrow_p\rangle,|\downarrow_e \uparrow_p\rangle,|\downarrow_e \downarrow_p\rangle\}$, where $|\uparrow_q\rangle$ and $|\downarrow_q\rangle$ are the electron ($q = e$) and proton ($q = p$) spin-up and spin-down states, $|\uparrow_e \uparrow_p\rangle$ is a state where  both the electron and proton spins are up, $|\uparrow_e \downarrow_p\rangle$ is a state where the electron spin is up, while the proton spin is down, $|\uparrow_e \uparrow_p\rangle=|\uparrow_e \rangle\otimes | \uparrow_p\rangle$, and $|\uparrow_e \downarrow_p\rangle=|\uparrow_e \rangle\otimes | \downarrow_p\rangle$. Each state ket in this four-dimensional space can thus be considered as a two-qubit state. 
\section{the concurrence}
We now consider a hydrogen atom in equilibrium with a heat reservoir at a temperature $\tau$. Pertinent to the hyperfine structure of an H atom at such thermal equilibrium is the density operator $\rho(\tau)=\frac{1}{Z}e^{-\beta {H_{HF}}}$, where $Z=Tr(e^{-\beta {H_{HF}}})$ is the partition function, $\beta=1/(k_B\tau)$, and $k_B$ is the Boltzmann constant. While quantum entanglement in pure quantum states is adequately understood in terms of the von Neumann entropy \cite{Bennett, Nielsen}, a suitable quantifier for the entanglement of mixed states is much harder to define. Yet, the entanglement of formation has been shown to satisfactorily address this problem  \cite{Wootters,Hill}, providing a closed-form computable measure for the entanglement of mixed states, fully applicable to thermal states as defined by the density operator $\rho(\tau)$ \cite{Arnesen, Werlang}. Specifically, given $\rho$ = $\rho(\tau)$, the entanglement of such states is meaningfully quantified via the concurrence \cite{Wootters,Hill} $ C=\textrm{max}\{0,\lambda_1-\lambda_2-\lambda_3-\lambda_4\}$, where $\lambda_i$, enumerated such that, $\lambda_1\geq\lambda_2\geq\lambda_3\geq\lambda_4$, are the eigenvalues of the Hermitian matrix $R=\sqrt{\sqrt{\rho}\tilde{\rho}\sqrt{\rho}}$ and $\tilde{\rho}=(\sigma_y\otimes\sigma_y)\rho^*(\sigma_y\otimes\sigma_y)$.

With $H_{HF}$ as defined by Eq. (1), the concurrence $C$ becomes

\begin{align}
  C= \frac{1}{\mathcal{G}}\textrm{max}\{0,\frac{\sinh[2\beta \mathcal{A} \sqrt{1+\xi^2}]}{\sqrt{1+\xi^2}}-e^{-2\beta \mathcal{A}}\},
   \end{align}
where, $\mathcal{G}=e^{-2\beta \mathcal{A}} \cosh(2\beta \mathcal{A} \xi)+\cosh[2\beta \mathcal{A} \sqrt{1+\xi^2}].$

In Fig. 1(b), we plot the concurrence $C$ calculated with the use of Eq. (3) as a function of temperature $T=1/(\beta \mathcal{A})$. Entanglement is seen to decrease with a rise in $T$, vanishing as the temperature is allowed to grow above a critical value $T_c$. When no magnetic field is applied, $\xi=0$, Eq. (3) yields $T_c$ = 4/ln3. It is instructive to rewrite this relation as $k_B \tau_c=\Delta E$/ln3, where $\Delta E$ = $E_{b,c,d} - E_a$ = $4\mathcal{A}$ is the energy gap between the highest- and lowest-energy HFS levels at $B$=0. To articulate the connection of the critical temperature $\tau_c$ to the fundamental physical constants, we express it as $$\tau_c=\frac{2}{3  \ln{3}}\frac{\alpha^2\hbar^2}{k_B a_0^2} \frac{g_e g_p}{m_p} =\frac{4\alpha^2\hbar c}{3 k_B \ln{3}} \frac{m_e}{m_p}g_e  g_p R_{\infty },$$
where $\alpha$ is the fine-structure constant and $R_{\infty }=m_{\text{e}}e^{4}/(8\varepsilon _{0}^{2}h^{3}c)$ is the Rydberg constant.

The HFS entanglement can thus survive only as long as the thermal energy $k_B \tau$ is much lower than the HFS splitting $\Delta E$. The maximum thermal energy that the natural HFS entanglement (i.e., the HFS entanglement at $B$ = 0) of the H atom can withstand is $k_B \tau_c\approx 5.35 \times 10^{-6}$ eV, or $\tau_c\approx 60$ mK. 

The effect of the magnetic field on the concurrence $C$ is much more complex and much less transparent. At very low $T$, where the natural HFS entanglement is high ($T \leq$ 3 in Fig. 1b), the concurrence $C$ is a rapidly decreasing function of $\xi$. However, as $T$ increases, approaching $T_c$, $C$($\xi$) gradually flattens out (Fig. 2a), eventually becoming nonmonotonic (for $T \geq$ 3.5 in Fig. 2a). This behavior of $C$($\xi$) is in agreement with the earlier observations by Zhu et al. \cite{Zhu}.

It is, however, above the $\tau_c$ threshold that the magnetic field can have the most dramatic and in many ways counterintuitive effect on the HFS entanglement of the H atom. As one of the central findings of our analysis, which goes well beyond the scope of Ref. \cite{Zhu},  when applied to an H atom at temperatures $T$ $\succsim$ $T_c$, an external magnetic field can induce an entanglement of HFS states against all the odds of thermal effects, which become strong enough at these temperatures to completely suppress the entanglement of HFS states in the $B$ = 0 regime. 

This effect is illustrated in Fig. 2b. Here, the temperature $T$ is set above the $T_c$ threshold. Thus, as long as $B$ = 0, and, hence $\xi$ = 0, the H atom features no HFS entanglement, $C$ = 0 (see Fig. 1b). Yet, even though all the natural HFS entanglement has been lost at these temperatures, an external magnetic field can overcome this loss and induce an entanglement in the HFS manifold. This effect will be referred to hereinafter as magnetically induced entanglement (MIE). The highest MIE concurrences $C$ are achieved at temperatures right above $T_c$ (the solid curve in Fig. 2b, corresponding to $T$ = 4). At higher $T$, thermal effects, once again, take their toll, reducing the maximum attainable MIE. Higher $T$, as can be seen in Fig. 2b, require stronger magnetic fields for the onset of MIE. Finally, in the regime of very strong magnetic fields ($\xi >$ 25 in Fig. 2b), $C$($\xi$) is seen to display a temperature-independent asymptotic behavior, with $C$($\xi$) curves plotted for different $T$ bunching together as a part of this asymptotic behavior to a $T$-independent universal $C_0$($\xi$) curve.

  \begin{figure}
\includegraphics[width=\columnwidth]{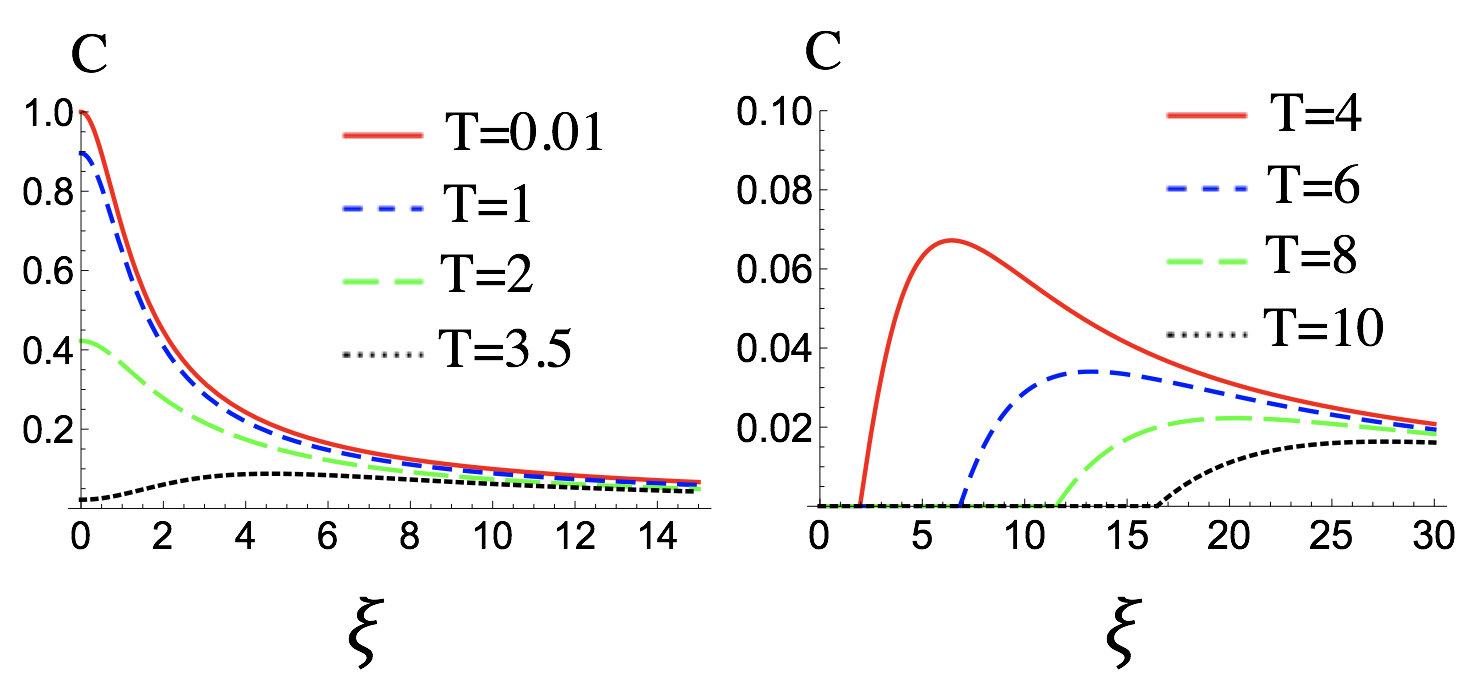}
\centering
\caption{ The concurrence of the HFS states of the H atom as a function of the normalized magnetic field $\xi$ below (a) and above (b) the critical temperature $T_c$.
}
\label{fig:2}
\end{figure}

To gain insights into the physics behind this behavior of the HFS entanglement as a function of the temperature and the magnetic field, we first examine the concurrence of the HFS states with no magnetic field applied, $B$ = 0. As one important finding, we see that, in this case, $ |c\rangle =\frac{1}{\sqrt{2}}(|\uparrow_e \downarrow_p\rangle+
  |\downarrow_e \uparrow_p\rangle)$
  and 
  $|a\rangle =\frac{1}{\sqrt{2}}(|\uparrow_e \downarrow_p\rangle-
  |\downarrow_e \uparrow_p\rangle)$ are maximally entangled, with their concurrence $C$ = 1. At low temperatures, such that $k_B \tau_c << \Delta E$ = $E_{b,c,d} - E_a$, the population of the lower-energy $|a\rangle$ singlet state is much higher than the population of the $|b\rangle$, $|c\rangle$, and $|d\rangle$ HFS triplet. As the temperature increases, however, the population of the HFS triplet grows, leading to a decrease in the HFS entanglement. Eventually, at $k_B \tau_c\approx \Delta E$, the population of the HFS triplet becomes comparable to the population of the $|a\rangle$ state, and the entanglement is completely lost, $C$ = 0. 
  
 An external magnetic field lifts the degeneracy of the $|b\rangle$, $|c\rangle$, and $|d\rangle$ states, splitting this triplet into three states with different, $B$-dependent energies (Fig. 1a). As the magnetic field increases, the energy gaps $\delta E_{c,d}$ = $E_{c,d} - E_a$ grow. It is straightforward to see from Eq. (3) that, in the low-temperature limit, $T <<$ 1, the temperature dependence of the concurrence is suppressed, $C$ $\approx$ $1/\sqrt{1+\xi^2}$, explaining the low-$T$ plateaus in the $C$($T$) plots in Fig. 1b.
 
 When the magnetic field is strong enough to induce $\delta E_{c,d}$ $\succsim$ $k_B \tau_c$, the temperature $\tau_c$ is no longer sufficient for complete entanglement suppression. More rigorously, as can be seen from Eq. (3), the HFS manifold retains entanglement as long as
  \begin{align}\label{Crit-m}
 {\sinh[2\beta \mathcal{A} \sqrt{1+\xi^2}]}-\sqrt{1+\xi^2}e^{-2\beta \mathcal{A}}> 0.
   \end{align}

 Thus, the critical magnetic field, $\xi_c$, needed to induce an HFS entanglement above the $\tau_c$ threshold can be found from  
\begin{align}\label{Crit-m}
 \sinh[2\beta \mathcal{A} \sqrt{1+\xi_c^2}]=e^{-2\beta \mathcal{A}}\sqrt{1+\xi_c^2}.
   \end{align}

In the $T >>$ 1 regime, high magnetic fields are required to induce HFS entanglement, $\xi_c>>1$. In this limit, Eq. (5) reduces to $2\beta \mathcal{A}\approx\ \ln{2\xi_c}/(\xi_c+1)$.

Specifically, with $\xi_c$ set at $\xi_c\approx16.5$, this approximation yields an estimate, $T\approx 10.01$, that agrees very well with the results of numerical simulations presented in Fig. 2b, where the dotted line, plotted for $T=10$, is seen to take off from the $C$ = 0 level at precisely   $\xi_c\approx16.5$.

\section{coherence}

As a useful insight, the inequality of Eq. (4), which guarantees the existence of HFS entanglement, can be expressed via the Manhattan-norm coherence, $D=\sum_{{i\neq j}} |\rho_{i,j}|$, also referred to as the $l_1$-norm coherence \cite{Baumgratz}. For the HFS states of the H atom at thermal equilibrium with a temperature $\tau$, this coherence is given by
\begin{align}\label{Crit-m}
 D= \frac{\sinh[2\beta \mathcal{A} \sqrt{1+\xi^2}]}{\mathcal{G}\sqrt{1+\xi^2}}.
   \end{align}

Eq. (4) is thus equivalent to $ D>e^{-2\beta \mathcal{A}}/\mathcal{G}$. Figures 3a and 3b illustrate the behavior of $D$ as a function of the temperature $T$ and the normalized magnetic field $\xi$. In the low-temperature limit, $T <<$ 1, $D$ is indistinguishable from the concurrence, $D$ $\approx$ $C$ $\approx$ $1/\sqrt{1+\xi^2}$. In this limit, the temperature dependence of the coherence is suppressed, leading to well-resolved plateaus in the $D$($T$) plots at low $T$ (Fig. 3a). 

In the limit of a strong magnetic field, $\xi >>$ 1, the coherence can be approximated as  $D\approx (1- e^{-2\beta \mathcal{A}})/\xi$. In this limit, the $D$($\xi$) plots bunch together (Fig. 3b) as $D$ tends to zero regardless of $T$. This behavior of $D$ in the $\xi >>$ 1 limit provides deeper insights into the properties of MIE at high $T$. Indeed, since the critical magnetic field $\xi_c$ needed to induce HFS entanglement can be found from $ D(\xi_c)=e^{-2\beta\mathcal{A}}/\mathcal{G}$, the maximum concurrence attainable in the MIE regime (Fig. 2b) is bound to follow a rapid roll-off of $D$($\xi$) as a function of $\xi$ (cf. Fig. 3b).

\section{THE PHYSICS BEHIND MAGNETICALLY INDUCED
ENTANGLEMENT OF H}

To better connect to the HFS physics behind the properties of MIE concurrence, it is instructive to examine Eq. (3) for $C$ jointly with the equations for the energy eigenvalues $E_u$. When no magnetic field is applied, $\xi$ = 0, the $|a\rangle$ and $|c\rangle$ states are maximally entangled, with their concurrence $C$ = 1. In the opposite limit of $\xi >> 1$, however, the $|a\rangle$ and $|c\rangle$ state kets tend to $|c\rangle \approx |\uparrow_e \downarrow_p\rangle$ and $|a\rangle \approx |\downarrow_e \uparrow_p\rangle$, with their energy eigenvalues given by $E_c \approx \mathcal{A}(-1+2\xi)$ and $E_a \approx \mathcal{A}(-1-2\xi)$. It is straightforward to see that the concurrence of these states tends to zero. Physically, this result is understood in terms of a strong coupling of electron and nuclear spins to the external magnetic field, which eventually breaks the spin--spin electron--nucleus coupling.

  \begin{figure}
\includegraphics[width=\columnwidth]{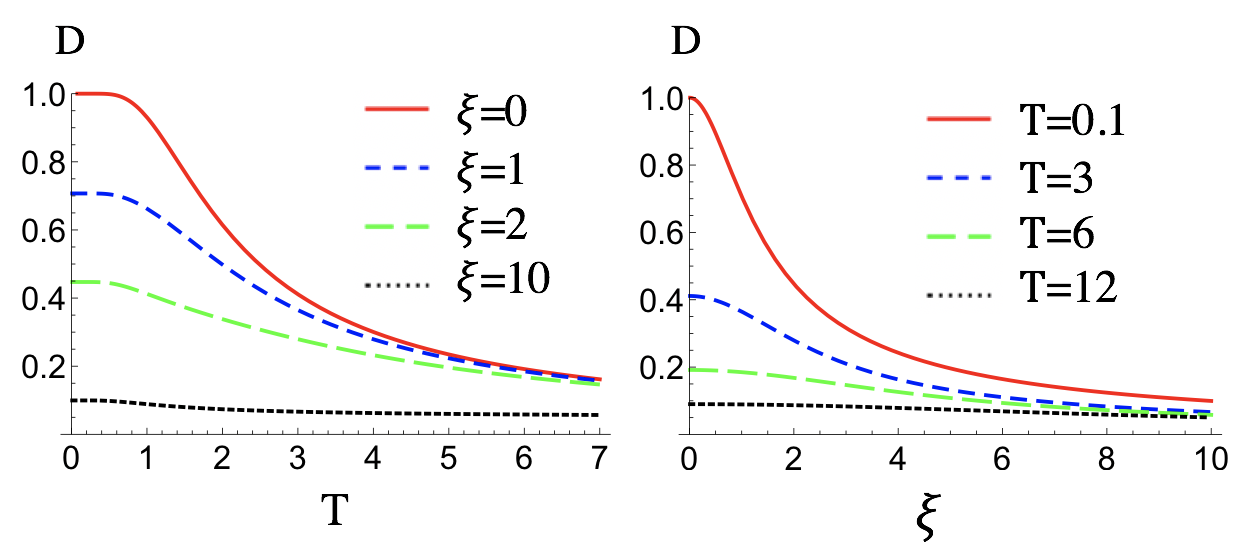}
\centering
\caption{ The $l_1$-norm coherence of the HFS states of the H atom as a function of the temperature $T$ (a) and the normalized magnetic field $\xi$ (b). 
}
\label{fig:4}
\end{figure}

We will now benchmark the HFS entanglement of the H atom against the two-qubit spin--spin entanglement in a one-dimensional (1D) Heisenberg chain (HC) \cite{Arnesen}. To this end, we consider two spins in a 1D Heisenberg chain driven by an external magnetic field $B$ governed by an interaction Hamiltonian
 \begin{align}\label{Hamilton}
   H_s=J\sigma_1\cdot\sigma_2+\mu_B B (\sigma_1^z+\sigma_2^z),
\end{align}
with a coupling constant $J <$ 0 ($J >$ 0) for a ferromagnetic (antiferromagnetic) system.

 Solving $ H_{s} |v\rangle=E_v  |v\rangle$, we find four energy eigenstates, $v =$ 1, 2, 3, 4, $|1\rangle=|\uparrow_1 \uparrow_2\rangle$,
   $|2\rangle=|\downarrow_1 \downarrow_2\rangle$,
   $|3\rangle =\frac{1}{\sqrt{2}}(|\uparrow_1 \downarrow_2\rangle+
  |\downarrow_1 \uparrow_2\rangle)$,
  $|4\rangle =\frac{1}{\sqrt{2}}(|\uparrow_1 \downarrow_2\rangle-
  |\downarrow_1 \uparrow_2\rangle)$, with energy eigenvalues $E_1 = J(1+4\xi)$, $E_2 =J(1-4\xi)$, $E_3 =J$, and $E_4 =-3J$, and the normalized magnetic field $\xi$ redefined as $\xi=\frac{ \mu_B B}{2J}$. 
  
  A two-qubit spin--spin entanglement in such a system is quantified in terms of the concurrence \cite{Arnesen}
\begin{equation}
  C= \textrm{max}\{0,\frac{e^{4\beta J}-3}{1+e^{4\beta J\xi}+e^{-4\beta J\xi}+e^{4\beta J}}
  \}.  
\end{equation}

The behavior of this concurrence as a function of the temperature $T$ and the normalized magnetic field $\xi$ is illustrated in Figs. 4a and 4b. As one of its central properties, readily seen from Eq. (8), $C$ = 0 for $k_B \tau_c>4J$/ln3 and $C =(e^{4\beta J}-3)/(1+e^{4\beta J\xi}+e^{-4\beta J\xi}+e^{4\beta J})$ for $k_B \tau_c<4J$/ln3.    

The spins thus remain entangled only for temperatures below $\tau_c=4J/(k_B$ln3).  

Similar to the HFS states of the H atom, the 1D Heisenberg chain is permissive of low-temperature spin--spin entanglement. The key properties of this entanglement, however, are profoundly different from the properties of HFS entanglement. The origins of these differences trace back to the physics of hyperfine splitting vis-à-vis the physics embodied by the 1D Heisenberg chain. Reflecting the difference in the physical contents of the two systems are the properties of the eigenfunctions $|u\rangle$ and $|v\rangle$ as dictated by the respective Hamiltonians $ H_{HF}$ and $ H_{s}$. When expanded in the basis a suitable for the description of spin--spin entanglement, viz., the $\mathcal{Q}(1,2)$ basis, the $|u\rangle$ eigenkets manifest a strong, explicit dependence on the magnetic field. By contrast, the $|v\rangle$ eigenkets in the same representation are $B$-independent. As a consequence, at $T$ = 0, the 1DHC concurrence is $C$ = 1 for any $\xi <$ 1 and $C$ = 0 for all $\xi >$ 1 (Fig. 4a). This property of $C$ is in stark contrast with the low-$T$ behavior of the HFS concurrence, which decreases monotonically with $\xi$ at any $T <<$ 1, $C$ $\approx$ $1/\sqrt{1+\xi^2}$ (Fig. 2a). 
When plotted as a function of the dimensionless magnetic field $\xi$ at $T\approx$ 0, the 1DHC concurrence displays a steep, almost stepwise change at $\xi$ = 1 (Fig. 4b). The HFS concurrence, on the other hand, is always a smooth, well-behaved function of $\xi$ (Figs. 2a, 2b). 

Because the magnetic field makes no imprint on the expansion coefficients of the $|v\rangle$ eigenkets in the $\mathcal{Q}(1,2)$ basis, the critical temperature for the 1DHC entanglement, $\tau_c=4J/(k_B$ln3), is independent of the magnetic field. Thus, unlike the natural HFS entanglement, whose critical temperature, i.e., the temperature at which $C$ = 0, is a function of $\xi$ (Fig. 1b), the 1DHC entanglement vanishes at the same temperature regardless of $\xi$ (Fig. 4a). Importantly, once it is gone with $\tau$ increased above $\tau_c$, the 1DHC entanglement never comes back no matter how strong the external magnetic field is. 

In this regard, the magnetically induced entanglement of HFS states of the H atom identified in this work is fundamentally different, as it is induced at temperatures above the critical point $\tau_c$, rising from zero (Fig. 2b). Once induced by a magnetic field with a critical magnitude $\xi_c$, the MIE of HFS states grows with $\xi$, reaches its temperature-dependent maximum, then decreases and eventually vanishes (Fig. 2b) as the magnetic field becomes strong enough to break the spin--spin coupling of the electron and its nucleus in the H atom.

\section{conclusion}
To summarize, we have shown that the electron and nuclear spin degrees of freedom in a hydrogen atom provide a physically meaningful frame and lend a suitable Hilbert space for bipartite entanglement, whose two-qubit concurrence and quantum coherence can be expressed in terms of the fundamental physical constant. The intrinsic, natural entanglement that the hyperfine-structure states of the H atom store at low temperatures is shown to rapidly decrease with a growth in temperature, vanishing above a critical temperature of $T_c$ $\approx$ 5.35 $\mu$eV. An external magnetic field, however, can overcome this thermal loss of HFS entanglement. As one of the central findings of this work, an external magnetic field can induce and sustain the entanglement of HFS states, against all the odds of thermal effects, at temperatures well above the  $\tau_c$ threshold, thus enabling magnetic-field-assisted entanglement engineering in low-temperature gases and solids. Experiments to isolate spectroscopic signatures of entanglement in hydrogen and hydrogen-containing systems are currently in progress as a follow-up to the earlier studies  \cite{Sheludiakov,Ahokas,Ahokas2}.

This research was supported in part by the Russian Foundation for Basic Research (projects Nos. 18-29-20031, 19-02-00473), Russian Science Foundation (project No. 20-12-00088 -- ultrabroadband optical science), Ministry of Science and Higher Education of the Russian Federation (project 14.Z50.31.0040, Feb. 17, 2017), and the Welch Foundation (grant A-1801-20180324).

  \begin{figure}
\includegraphics[width=\columnwidth]{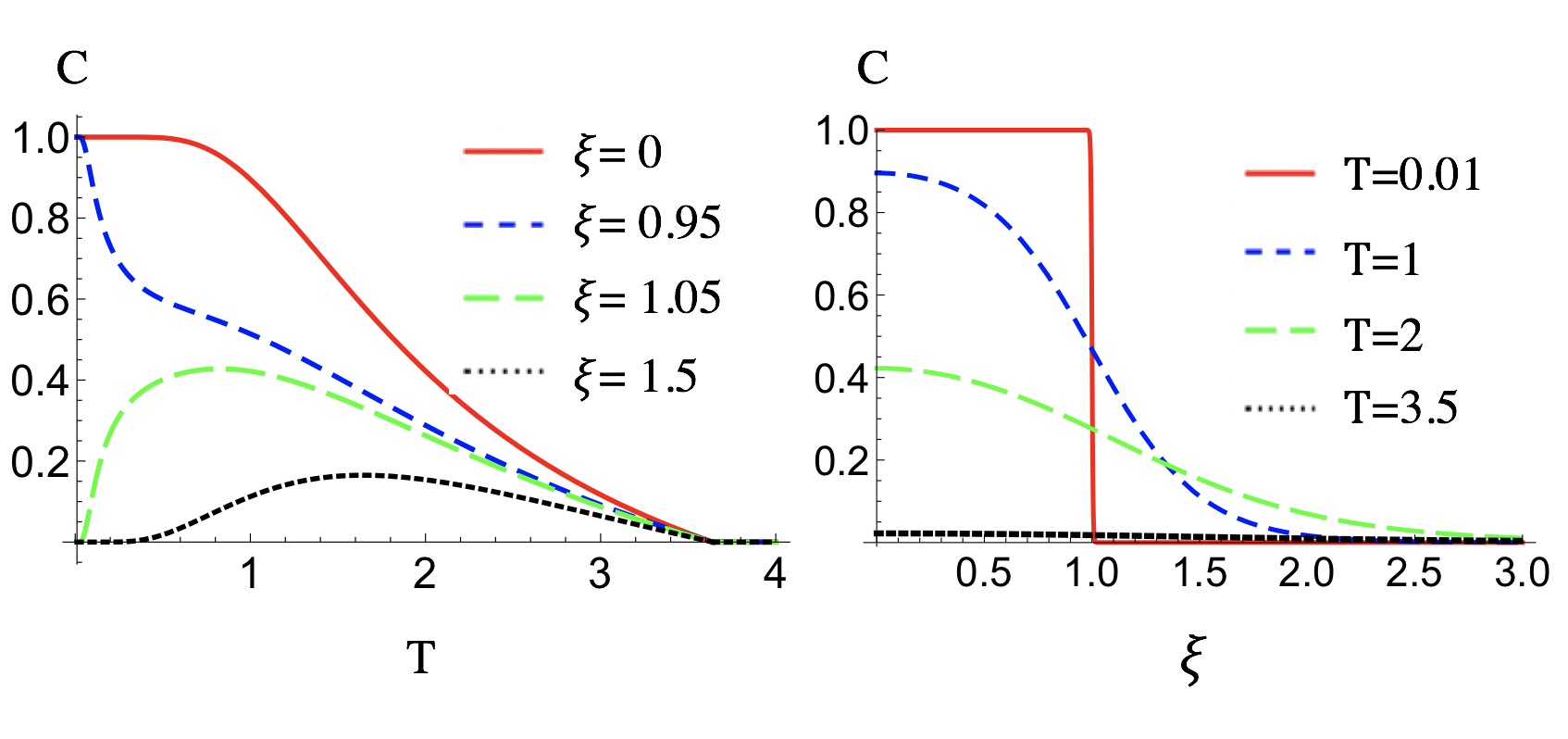}
\centering
\caption{The concurrence $C$ quantifying the two-qubit entanglement in a 1D Heisenberg model as a function of the temperature $T$ (a) and the normalized magnetic field $\xi$ (b).
}
\label{fig:s1}
\end{figure}

\end{document}